\documentclass[a4paper]{jpconf}
\usepackage{graphicx}
\begin{document}
\title{Proposal for a magnetic field induced graphene dot}

\author{P. A. Maksym$^1$, M. Roy$^1$, M. F. Craciun$^2$
  \footnote[4]{Present Address: School of Physics, University of Exeter,
  Exeter EX4 4QL, UK}, S. Russo$^{2\ 4}$,
  M. Yamamoto$^2$, S. Tarucha$^2$ and H. Aoki$^3$}

\address{$^1$Department of Physics and Astronomy, University of Leicester,
  Leicester LE1~7RH, UK}

\address{$^2$Department of Applied Physics, University of Tokyo,
Hongo, Bunkyo-ku, Tokyo 113-8656, Japan}

\address{$^3$Department of Physics, University of Tokyo,
Hongo, Bunkyo-ku, Tokyo 113-0033, Japan}

\begin{abstract}
Quantum dots induced by a strong magnetic field applied to a single layer of
graphene in the perpendicular direction are investigated.
The dot is defined by a model potential which consists
of a well of depth $\Delta V$ relative to a flat asymptotic part and quantum
states formed from the zeroth Landau level are considered. The energy of
the dot states cannot be lower than $-\Delta V$ relative to the asymptotic
potential. Consequently, when $\Delta V$ is chosen to be about half of the
gap between the zeroth and first Landau levels, the dot states are isolated
energetically in the gap between Landau level 0 and Landau level -1. This
is confirmed with numerical calculations of the magnetic field dependent
energy spectrum and the quantum states. Remarkably, an antidot formed by
reversing the sign of $\Delta V$ also confines electrons but in the energy
region between Landau level 0 and Landau level +1. This unusual behaviour
gives an unambiguous signal of the novel physics of graphene quantum dots.
\end{abstract}

\section{Introduction}
Graphene is an astonishing material with a zero energy gap and massless
charge carriers. This allows relativistic physics to be studied in the
solid state and leads to many remarkable phenomena \cite{Geim07}. One of
the most relevant to graphene dots is the quantum Klein paradox. This makes
all potential barriers perfectly transparent to graphene charge carriers
moving in one and two dimensions \cite{Katsnelson06} so it is
impossible to confine graphene carriers in an external potential well.
However it is possible to confine them with a
combination of electric and magnetic fields. Essentially, the magnetic
field deflects the carriers so that they move in closed orbits and remain
localised but this kind of confinement cannot be achieved with an arbitrary
combination of electric and magnetic fields. The conditions needed have
been analysed by Giavaras, Maksym and Roy
\cite{Giavaras09} who have suggested a device structure that can be used to
probe both confined and deconfined states in graphene dots. The present
work is about a simpler structure that can be used to focus on confined
states.

Conditional confinement is the one of the most important differences
between graphene and conventional semiconductors. It is a direct
consquence of the lack of mass. Roughly speaking, confined quantum states
occur only when all the corresponding classical orbits are bounded but when
massless particles move in a weak magnetic field \textit{both} bounded
\textit{and} unbounded orbits occur at the same energy. Raising the
magnetic field removes the unbounded orbits and enables confined states to
occur. Details are given in ref. \cite{Giavaras09} for circularly symmetric
scalar potentials, $V = V_0 r^s$ and azimuthal vector potentials,
$A_{\theta} = A_0r^t$, where $r$ is the radial distance and $s, t > 0$.  In
this case all the states are confined when $t>s$ but when $t=s$ the states
are confined only when $c^2e^2A_0^2 > V_0^2$, where $c$ is the Fermi
velocity of the graphene sheet. Since $V_0$ and $A_0$ are experimentally
tunable parameters this allows both confined and deconfined states to be
probed in graphene dots and the interesting transition between the two
regimes that occurs when $c^2e^2A_0^2 = V_0^2$ could also be probed
experimentally.

Another important difference between graphene and conventional
semiconductors is that graphene is gapless. Dots in conventional
semiconductors are engineered so
that the dot states are both confined and have energies in the band
gap. Thus the dot states are spatially and energetically isolated from the
rest of the system. The dot is an artificial atom and any probe of energy
levels in the gap, for example charging experiments, is only sensitive to dot
states and does not respond to bulk states. It is more difficult to achieve
this in graphene because there is no natural gap. One option is to engineer
a device structure that puts confined dot states in the gap between bulk
Landau levels. However there is a problem. The potentials in a real
graphene dot cannot rise indefinitely as a power law and must become flat
asymptotically. So there are bulk Landau states outside the dot and it is
necessary to ensure that the confined dot states occur in a region of low
density of states away from the broadened bulk Landau levels. Giavaras,
Maksym and Roy \cite{Giavaras09} showed that it is possible to achieve this
while allowing the confinement-deconfinement transition to be observed
provided that the scalar potential consists of a well which is separated
from the asymptotic region by a barrier. However the barrier is not needed
if one only wants to probe confined states. Then one can use a simple
potential well as shown in figure \ref{ebfigure}.

A well in a uniform, perpendicular magnetic field clearly satisfies the
confinement condition when the field is strong enough. The reason why it
also satisfies the condition for energetic isolation is related
to the unusual physics of the zeroth Landau level in graphene. In brief,
the zeroth Landau level has zero energy and when a potential well
of depth $\Delta V$ is applied to graphene the energy of states formed from
the zeroth Landau level does not shift by more than about $|\Delta V|$. So
dot states can be isolated energetically by choosing $\Delta V$ to put them
into the middle of the Landau gap. The same result applies to antidot
states, which are also confined because of the symmetry of the Dirac
cone. So observations of the symmetry between dot and antidot confined
states would give a direct experimental confirmation of the influence of
the Dirac cone on the graphene dot energy spectrum. This physics is
detailed in section \ref{PhysicsSection} and numerical studies of the dot
energy spectrum and quantum states are reported in section
\ref{NumericalSection} where the properties of the graphene dot are
compared with those of a typical electrostatic dot formed from
GaAs.

\section{Physics of Electron Confinement in the Zeroth Landau Level}
\label{PhysicsSection}

The graphene sheet is taken to be in a uniform, perpendicular magnetic
field, $B$ and the azimuthal component of the magnetic vector potential at
position $\mathbf{r}$ is $A_\theta = Br/2$, where $\theta$ is the azimuthal
angle. The scalar potential is taken to be
$V(r) = \pm(V_0\cos(\pi r/2R) + V_1), r < 2R$,\quad
$V(r) = \pm(V_1 - V_0),r \ge 2R$,\quad $V_0, V_1 > 0$,
where the $+$ sign corresponds to an antidot, the $-$ sign corresponds to
a dot and $R$ is a size parameter. In terms of these parameters the depth
of the dot and height of the antidot are given by $|\Delta V| =
2V_0$. Because the solution to the Laplace equation can be written as a
Fourier series, this simple form is a crude model of the potential in a real
dot. 

The potentials have circular symmetry so the quantum states are eigenstates
of angular momentum, $\hbar m$, where the angular momentum quantum number
is the integer $m$. The graphene states are two-component states where each
component gives the probability amplitude for an electron being on one of
the two sub-lattices of the graphene sheet. The
two-component envelope function is
$(\chi_1(r)\exp(i(m-1)\theta),\chi_2(r)\exp(i m\theta))$ and the radial
functions, $\chi_i$ satisfy \cite{Giavaras09}
\begin{eqnarray}
\frac{V}{\gamma}\chi_1 - i\frac{d\chi_2}{dr} -i \frac{m}{r}\chi_2 -
i\frac{e}{\hbar}A_{\theta}\chi_2 &=& \frac{E}{\gamma}\chi_1,\nonumber\\ 
- i\frac{d\chi_1}{dr} +i \frac{(m-1)}{r}\chi_1 +
i\frac{e}{\hbar}A_{\theta}\chi_1 + \frac{V}{\gamma}\chi_2 &=& 
\frac{E}{\gamma}\chi_2,\nonumber
\end{eqnarray}
where $\gamma=646$ meV nm and $E$ is the energy. The Fermi velocity is
given by $c=\gamma/\hbar$.

When $V=0$ these equations give the graphene Landau levels of energy
$E = \pm\gamma\sqrt{2N}/l$ where $l^2 = \hbar/eB$, $l$ is the magnetic length,
$N = n + (|m| + m)/2$ is the Landau quantum number and $n$ is a radial quantum
number. The zeroth Landau level states correspond to $n=0$ and $m\le
0$. Unlike the situation of a particle with mass, the energy of the zeroth
Landau level is exactly zero and there is no zero point energy. In addition
$\chi_1$ vanishes and $\chi_2 \propto r^{|m|}\exp(-r^2/4l^2)$ so that the
quantum states are localised on rings whose radius increases like
$\sqrt{|m|}$.

Now consider what happens to the zeroth Landau level states in the presence
of a non-zero potential. In this case $\chi_1$ no longer vanishes but it
can be shown that the zero point energy remains small
\cite{Giavaras10}. Consequently the energy of the states is determined
mainly by the local potential. In the case of a dot the lowest energy
states are those of smallest spatial extent so if the dot is large enough
that the $m=0$ state fits into it, the energy of the lowest state in the
dot is no more than $\Delta V$ below the asymptotic value of the potential.
When $m$ increases the states are localised in regions of weaker potential
so their energy tends to increase with $m$. In the limit of very large $m$,
the states merge into the bulk zeroth Landau level which is shifted from
zero by the asymptotic value of the potential. These statements can be proved
rigorously but the details are quite complicated \cite{Giavaras10}.
This physics is the key to isolating a confined dot state energetically. If
$\Delta V$ is chosen to be about half of the gap between Landau level 0 and
Landau level -1, that is $\Delta V \sim \gamma/l\sqrt{2}$, the lowest dot
state will be in the desired region of very low density of states.

To analyse the case of an antidot consider what happens when $V$ is
replaced by $-V$. It can be shown by direct substitution into the
radial equations that if
$(\chi_1(r)\exp(i(m-1)\theta),\chi_2(r)\exp(i m\theta))$ is an eigenstate
of energy $E$ corresponding to potential $V$, then
$(\chi_1(r)\exp(i(m-1)\theta),-\chi_2(r)\exp(i m\theta))$ is an eigenstate
of energy $-E$ corresponding to potential $-V$. Hence the antidot states
are confined whenever the dot states are confined and the energy of the
\textit{highest} antidot state is no more than $\Delta V$ about the
asymptotic value of the antidot potential. The fact that both dots and
antidots confine electrons is unique to graphene and is an observable
consequence of its Dirac cone band structure. 

\section{Numerical Studies of Dots and Antidots}
\label{NumericalSection}

\begin{figure}
\begin{center}
\includegraphics[width=3.3cm,angle=-90]{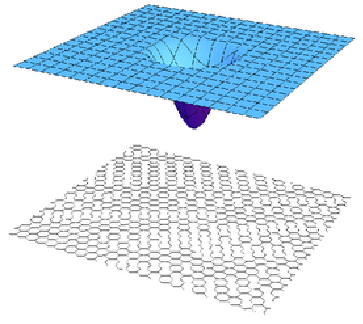}\quad 
\includegraphics[width=3.6cm,angle=-90]{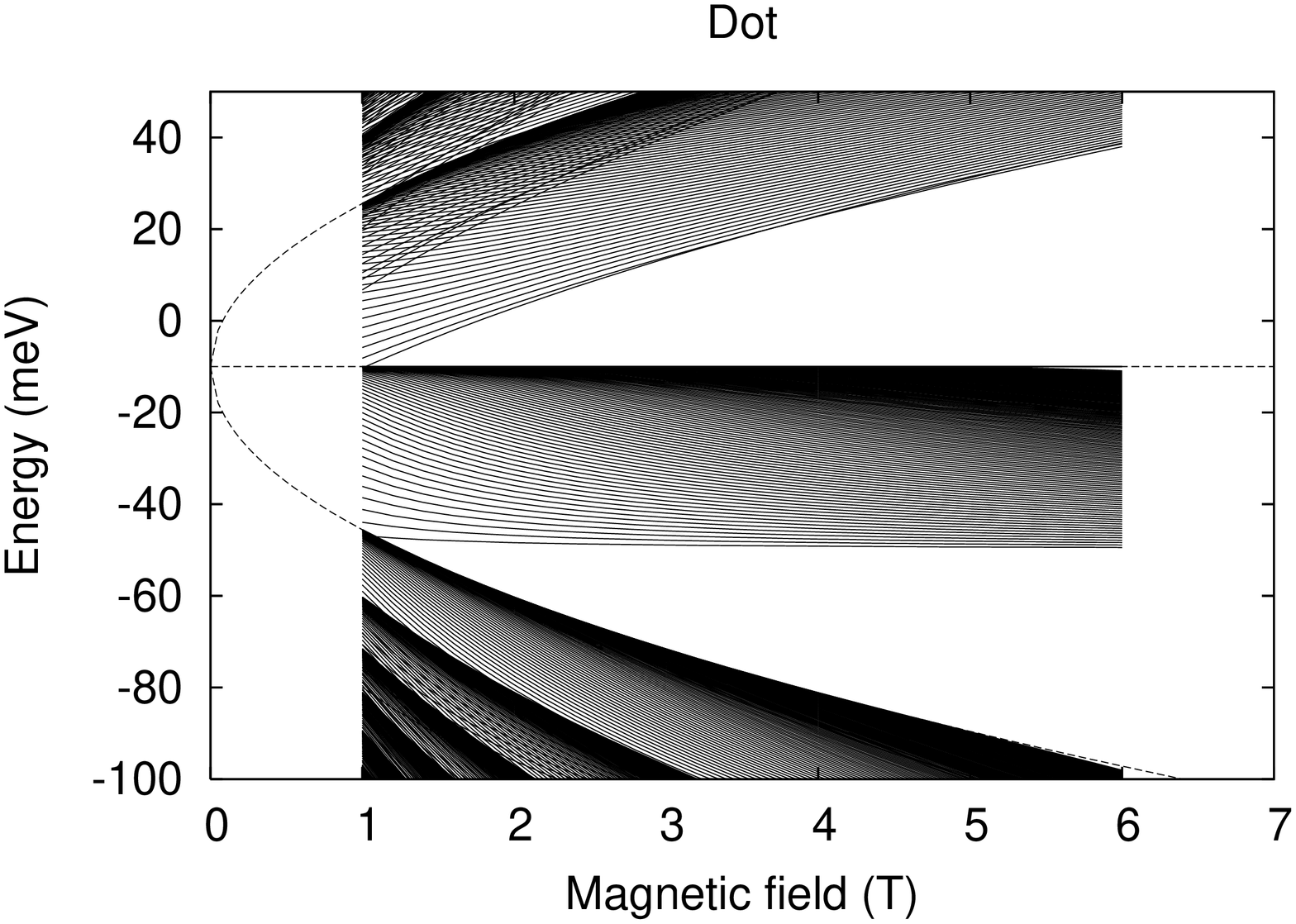}\quad 
\includegraphics[width=3.6cm,angle=-90]{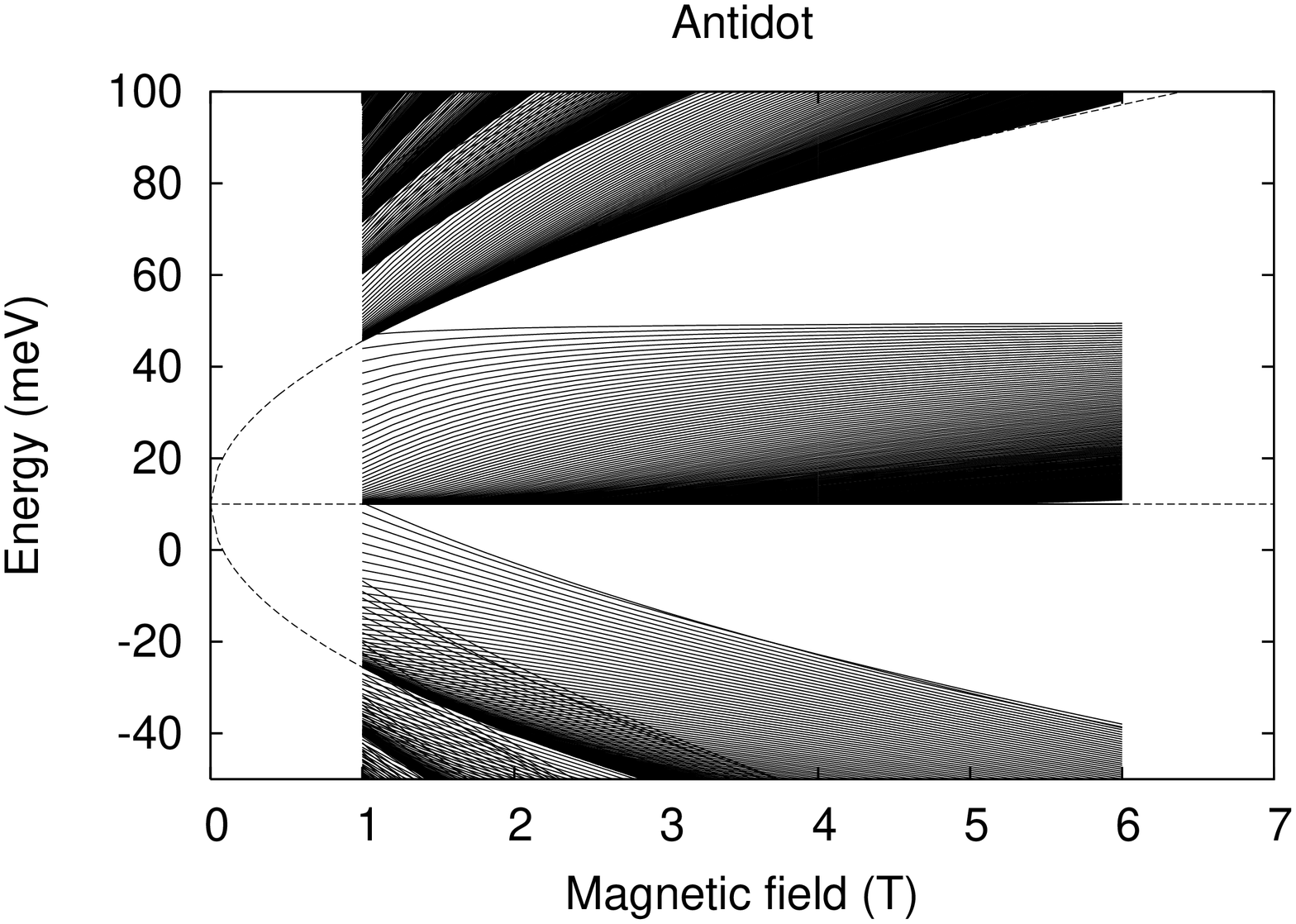}
\end{center}
\vspace{-0.3cm}
\caption{Physical model (left) and energy spectra of a dot (centre) and
  antidot (right). }
\label{ebfigure}
\end{figure}

\begin{figure}[h]
\begin{center}
\includegraphics[width=4.3cm,angle=-90]{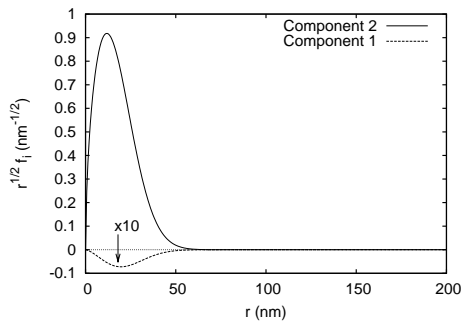}\quad%
\begin{minipage}[t]{5cm}
\vspace{2cm}
\caption{\label{statefigure}Two component radial envelope function for the
  $m=0$ state in a dot.}
\end{minipage}
\end{center}
\end{figure}

To compute energies and states, the radial equations are made real by putting
$f_1 = \chi_1$, $if_2 = \chi_2$, then $f_1$ and $f_2$ are found with the
method outlined in ref. \cite{Giavaras09}. The model parameters are
$V_0=20$ meV, $V_1=30$ meV and $R=100$ nm. Hence $\Delta V = 40$ meV, about
half the Landau gap at 5 T and
$V(r)$ becomes $\pm 10$ meV outside the dot or antidot.
Magnetic field dependent energy spectra are shown in
figure \ref{ebfigure}. Each spectrum consists of a superposition of energy
levels and the $m$ range, $-150\le m \le 10$, is chosen
so that all states localised in the dot or antidot are included in the field
range between 1 and 5.5 T. For comparison, the dashed lines show the Landau
levels of an ideal graphene sheet. These levels are shifted
by the asymptotic value of the potential, -10 meV in the case of the dot
and +10 meV in the case of the antidot. The effect of the external
potential is clearly to broaden the Landau levels. In the case of the dot,
the zeroth Landau level at -10 meV is broadened into a band that occupies
the region between -10 and -50 meV and the width of the band is very
insensitive to the magnetic field. As expected the width of the band
simply reflects the range of potential values: no zeroth Landau level
state rises in energy above -10 meV and no zeroth Landau level state falls
in energy below about -50 meV. In contrast, the upper edge of the broadened
$N=-1$ Landau level falls in energy like $-\sqrt{B}$ and the lower edge of
the broadened $N=+1$ Landau level rises in energy like
$+\sqrt{B}$. Therefore the band of $N=0$ dot states becomes energetically
isolated as the field increases. With the present model parameters, the
lowest dot state appears about in the middle of the gap when the field is
about 5 T. In a real graphene sheet the Landau levels are broadened because
of imperfections but the same principle applies and the 
magnetic field and potential can be chosen so that the lowest dot state
lies in a region of extremely low density of states between
broadened Landau levels. The right hand side of figure
\ref{ebfigure} shows the antidot spectrum. It has the same features as the
dot spectrum except that the band of confined antidot states appears in the
gap between the $N=0$ and $N=1$ Landau levels. The symmetry between the dot
and antidot spectra is clearly visible. Figure \ref{statefigure} shows the
lowest ($m=0$) state in a dot and demonstrates that it is confined within
the dot.

\vspace{-0.2cm}

\vspace{-0.8cm}
In summary, the proposed graphene dot is able to confine electrons in a
region that is
isolated both spatially and energetically. The typical spacing between the
dot levels at $B=5$ T ranges from $\sim 0.6$ meV for $R=100$ nm to
$\sim 2$ meV for $R=50$ nm. These sizes and level spacings are comparable
to those for a
GaAs electrostatic dot with confinement energy 4 meV in a magnetic field
range of 3-10 T. The energy levels in the GaAs dot can be resolved by
charging experiments, for example, so similar experiments on the graphene
dot should be feasible.

\ack
This work was supported by the UK Royal Society. PAM is grateful for study
leave from the University of Leicester and hospitality at the Department of
Physics, University of Tokyo where part of this work was done. ST and MY
acknowledge support from the Strategic International Cooperative Program
(Joint Research Type), Japan Science and Technology Agency.

\end{document}